\documentclass{article}
\usepackage[paperwidth=8.5in, paperheight=11in, margin=1in]{geometry}

\usepackage{algorithm}
\usepackage{algpseudocode}

\usepackage{array} 
\usepackage{comment} 
\usepackage{afterpage}

\usepackage[geometry]{ifsym}

\usepackage{url}                       
\usepackage{hyperref}                 
\usepackage{bm}

\usepackage{amssymb, amsmath, amsfonts, latexsym}
\usepackage[mathscr]{euscript}

\usepackage{mathtools}

\usepackage{fancyref}

\usepackage{rotating}

\usepackage{multirow}
\usepackage{array, makecell, rotating}

\def\indep{\perp\!\!\!\perp}

\font\Bigmath=cmsy10 scaled \magstep2

\def\diamondplustwo{\mathrel{%
  \ooalign{$+$\cr\hss\lower.255ex\hbox{\Bigmath\char5}\hss}}}

\newcommand{\calE}{\mathcal{E}}

\usepackage{tikz}
\usetikzlibrary{arrows,decorations.pathmorphing,backgrounds,positioning,fit,petri}

\usepackage{tabu} 

\usepackage[square,sort,comma,numbers]{natbib}

\makeatletter \renewcommand\@biblabel[1]{#1.} \makeatother

\begin{document}

\begin{titlepage}
\begin{center}
  \bfseries
  \huge Person as Population: A Longitudinal View of Single-Subject Causal Inference for Analyzing Self-Tracked Health Data
  \vskip.2in
  \textsc{\LARGE E. J. Daza, DrPH, MPS} \\
  \textsc{\LARGE \today}
  \vskip.2in
  \large {\sl Please send correspondence to:}
  \large Eric J. Daza \\
  \large ericjdaza@gmail.com
\end{center}
%
%
%
%
\end{titlepage}

\clearpage

\section*{Abstract}

Single-subject health data are becoming increasingly available thanks to advances in self-tracking technology (e.g., wearable devices, mobile apps, sensors, implants). Many users and health caregivers seek to use such observational time series data to recommend changing health practices in order to achieve desired health outcomes. However, there are few available causal inference approaches that are flexible enough to analyze such idiographic data. We develop a recently introduced causal-analysis framework based on n-of-1 randomized trials, and implement a flexible random-forests g-formula approach to estimating a recurring individualized effect called the \lq\lq average period treatment effect". In the process, we argue that our approach essentially resembles that of a longitudinal study by partitioning a single time series into periods taking on binary treatment levels. We analyze six years of the author’s own self-tracked physical activity and weight data to demonstrate our approach, and compare the results of our analysis to one that does not properly account for confounding.

\medskip
\noindent
{\bf Keywords:} causal inference, n-of-1 trial, single subject, time series, longitudinal, average period treatment effect

\clearpage

\section{Introduction}\label{sec:intro}

A longitudinal study can be used to estimate or infer mean trends over time under different values of an exposure, taken across a set of individual trends. Its target population consists of some larger set of individuals within a population of people. An n-of-1 or single-subject study can be used to estimate or infer mean trends over time under different values of an exposure, taken across a set of period-long trends where a period is an interval of time. Its target population consists of some larger set of time periods within a person. In this sense, an n-of-1 study can be treated as a sample from a population comprised entirely of events in one person's life: You are a world unto yourself, or {\sl mundong sarili mo} (Tagalog). In essence,\lq\lq you" are a \lq\lq y'all".

Single-subject, single-case, or {\sl n-of-1} studies refer to research focused on one individual. (In a {\sl self-tracked} study, the individual has recorded or measured her own data.) In psychology, these studies are called idiographic (i.e., population-of-one), while nomothetic (i.e., population-of-many) studies characterize groups of individuals \cite{2005_ponterotto}. Biomedical research and clinical trials in particular have employed n-of-1 studies \cite{1986_guyatt_etal, 1990_guyatt_etal, 1999_backman_harris, 2011_gabler_etal, 2011_lillie_etal, 2013_duan_etal}, with trial design, implementation, and analysis guidance available in various texts \cite{2014_naughton_johnston, 2014_kravitz_etal, 2015_nikles_mitchell, 2016_shamseer_etal, 2016_vohra_etal}. Chen et al (2012)\cite{2012_chen_etal} argued that wearable devices can be used to facilitate n-of-1 trials. And both a recent article in Nature \cite{2015_schork} and the U.S. Department of Health and Human Services Agency for Healthcare Research and Quality (AHRQ)\cite{2014_kravitz_etal} have considered n-of-1 trials to be part of \lq\lq personalized medicine".

N-of-1 randomized trials (N1RTs) do not generally require causal inference methods due to their simple randomized crossover design. However, it is natural to want to apply such methods to conduct causal inference in n-of-1 observational studies (N1OSs), wherein confounding and selection bias may exist. An N1OS is defined as a non-randomized single-subject study, in our case, with the objective of discovering causal effects of putative treatments that can be tested in an N1RT. Daza (2018)\cite{2018_daza} constructed a framework to analyze causal effects in a N1OS (and N1RT as the special randomized case) using the Neyman-Rubin-Holland counterfactual approach \cite{1923_neyman, 1974_rubin, 1986_holland}. The author demonstrated how to use this framework by applying two common causal inference methods (i.e., g-formula or regression adjustment, and propensity-score inverse probability weighting) to characterize the effect of his physical activity (PA) patterns on his weight, using data self-tracked over a six-year period. However, the author did not explicitly acknowledge the conceptual similarities between n-of-1 and longitudinal studies.

Causal inference methods have been extensively developed and deployed to analyze time-varying effects in observational longitudinal health data. These include marginal structural models and the time-varying g-formula \cite{2000_robins, 2000_robins_etal, 2009_robins_hernan}. While these methods can and should be extended to the n-of-1 case, they are not as easily deployed with current programming solutions. We therefore develop and apply a practical approach using the readily available randomForest package in R \cite{2002_liaw_wiener}. Specifically, we adapted the single-tree approach of Athey and Imbens (2015)\cite{2015_athey_imbens} to the n-of-1 case using random forests \cite{2001_breiman}. Tree-based techniques have the advantage over linearized approaches (e.g., elastic net) of being able to target a causal contrast in outcomes between levels of the exposure, while flexibly modeling the effects and associations of other causes, confounders, or covariates that are not of central interest.

Balandat (2016)\cite{2016_balandat} took a similar approach towards causal prediction of electricity consumption, and developed the large-sample theory needed to conduct inference on what Daza called the {\sl average period treatment effect} (APTE) \cite{2018_daza} if exposure effects do not carry over from one period to the next. Balandat demonstrated his approach using a prognostic-score-type technique \cite{2008_hansen} wherein the model fitted to control-group outcomes predicted by the other covariates was used to predict counterfactual control outcomes in the treatment group. Our single-tree approach instead includes the exposure itself as a predictor, and is a refinement and extension of the analysis performed by Cheung et al (2017)\cite{2017_cheung_etal}, whose work sets the stage for comparing nomothetic and idiographic modeling of the APTE using tree-based approaches. (Burg et al [2017] \cite{2017_burg_etal} accomplished a similar goal through mixed-effects modeling using generalized linear models. Balandat also drew a formal connection between individual- and group-level average treatment effects in his Section 4.5.)

Our goal in this paper is to demonstrate how n-of-1 causal effects can be reasonably specified and estimated in a longitudinal context. The rest of this paper is organized as follows. In Section \ref{sec:methods}, we define key notation and causal inference concepts for modeling and estimating the APTE using a counterfactual-based framework rooted in Daza (2018)\cite{2018_daza}, and used random forests to model the key component of the g-formula approach. We then estimate the APTE of the author's PA level on his weight in Section \ref{sec:emp.example} using the same six years of self-tracked data analyzed in \cite{2018_daza}. We conclude in Section \ref{sec:discuss} with a short summary and very brief proposal for how a set or series of n-of-1 studies can be aggregated to estimate a population-level average treatment effect. All analyses were conducted in R version 3.5.1.

\section{Methods}\label{sec:methods}


	\subsection{Notation, Definitions, and Assumptions}\label{subsec:notass}

In this section, we refine the counterfactual or potential-outcomes framework developed in Daza (2018)\cite{2018_daza}. Throughout this article, we use the following notation. Random variables and fixed values are written in upper-case and lower-case, respectively. Let $p ( A = a )$ denote the probability mass or density of random variable $A$ at $a$, with shorthand $p ( a )$. Let $\{ ( A ) \}$ denote a stochastic process; i.e., a time series of random variables. For any index $j$, let $\{ ( j ) \}$ denote a sequence. For any random variable $B$, let $B | A $ denote the relationship \lq\lq $B$ conditional on $A$", with shorthand $B | a $ for $B | A = a$. Let $B \indep A$ denote statistical independence of $B$ and $A$. Let $E ( \cdot )$ denote the expectation function.

Let $Y$ represent an outcome event of interest (hereafter, outcome), $X$ represent a categorically valued exposure event of interest (hereafter, exposure) that causes and therefore precedes $Y$, and $U$ represent the set of all other possibly unobserved events that both precede $X$ and cause $Y$. Let $g^Y \big( U, X, \calE^Y \big)$ denote the outcome {\sl causal mechanism} (CM); i.e., the event-generating mechanism for observed outcomes. (For continuous variables, we will assume the relevant error term $\calE$ is zero-mean with finite variance; for categorical variables, that $\calE$ is uniformly distributed.)

A potential-outcomes framework can be used to further specify $g^Y \big( U, X, \calE^Y \big)$ as follows. We define $Y^a$ to be a value of $Y$ corresponding to exposure level $a$ as a {\sl potential outcome} (PO). Specifically, we will call $Y^a = g^Y_a \big( U, \calE^Y \big)$ the PO-generating mechanism (POGM), where the POGM is indexed by $a$; i.e., as an index, $a$ is a fixed value that is neither an exposure nor cause of $Y$. The realized or observed outcomes are related to the POs by the equivalence $Y = \sum_a Y^a I ( X = a )$ for all $a$ in the support of $X$, where $I(b) = 1$ if expression $b$ is true and $I(b) = 0$ otherwise. This equivalence is called {\sl causal consistency} (CC), and ensures that the outcome we observe is identical to its corresponding PO. We can use CC and the POGM together to specify the outcome CM as $g^Y \big( U, X, \calE^Y \big) = \sum_a g^Y_a \big( U, \calE^Y \big) I ( X = a )$; here, the POGM is a functional component of the outcome CM. This specification formalizes the {\sl fundamental problem of causal inference} \cite{1986_holland} that only one PO (i.e., corresponding to $X = a$) can be observed for any given set of values of $U$ and $\calE$. Hence, the term \lq\lq counterfactual outcome" is also used to refer to $Y^{a'}$ for some $a' \ne a$ because if $X = a$ is in fact observed, then observation of $Y^{a'}$ runs \lq\lq counter to fact" (i.e., $Y^{a'}$ cannot be observed under CC).

For example, suppose $g^Y \big( U, X, \calE^Y \big) = \beta_0 + \beta_1 X + \beta_2 U ( 1 - X ) + \calE^Y$ and that $X \in \{ 0, 1 \}$ is a binary exposure. One natural set of the two corresponding POGMs would be $g^Y_0 \big( U, \calE^Y \big) = \beta_0 + \beta_2 U + \calE^Y$ and $g^Y_1 \big( U, \calE^Y \big) = g^Y_1 \big( \calE^Y \big) = \beta_0 + \beta_1 + \calE^Y$. Under CC, it clearly follows that $\beta_0 + \beta_1 X + \beta_2 U ( 1 - X ) + \calE^Y = ( \beta_0 + \beta_2 U + \calE^Y ) I ( X = 0 ) + ( \beta_0 + \beta_1 + \calE^Y ) I ( X = 1 )$. This example highlights the fact that the PO concepts we have defined (i.e., the CC-POGM formulas) are generally not required to investigate causality in terms of outcome CMs. We elaborate on this point in Section \ref{subsec:ate}, where we nonetheless assert that such a PO-based approach encourages development and implementation of pragmatic, intuitively discrete interventions.

Let $R = 1$ denote randomization (i.e., random selection or assignment) of $X$, and let $R = 0$ denote the absence of randomization (i.e., corresponding with the ecological, natural, or otherwise undisturbed state of $X$). Suppose $X$ is only caused by $U$ or $R$ (and random error) such that the exposure CM is $g^X \big( U, R, \calE^X \big)$. By standard definitions, randomization eliminates the dependence of $X$ on $U$, which we formalize as $g^X \big( U, R, \calE^X \big) = g^X_{R=1} \big( \calE^X \big) R + g^X_{R=0} \big( U, \calE^X \big) ( 1 - R )$. We will call this specification the exposure-randomization CM (ERCM). (Note that the {\sl do-operator} of Pearl (2009) \cite{2009_pearl} can be considered shorthand notation for \lq\lq X, R=1"; in particular, $p \big( y | do(x) \big) = p ( y | x, R = 1 )$ and $p \big( y | x \big) = p ( y | x, R = 0 )$.)

Statistical causal inference commonly involves three conditions. The first two conditions follow if the POGM and ERCM hold. When the distributions of potential outcomes do not depend on the assignment mechanism, we say that the assignment mechanism is ignorable and that {\sl ignorability} (equivalently, {\sl exchangeability}) holds. This is implied by the ERCM; i.e., $\big\{ Y^a \big\} \indep X \big| R = 1$. If this holds in general given all other causes, we call this condition {\sl conditional ignorability/exchangeability}; i.e., $\big\{ Y^a \big\} \indep X | U$. The last condition, called {\sl positivity} or {\sl overlap}, is the empirical requirement that each exposure level must sufficiently co-occur with all other causes. For example, if $U \in \{ 0, 1 \}$, then positivity holds if there is at least one observation with $U = 0$ and one with $U = 1$ for each level of $X$ observed. Positivity is required in order to perform estimation.

	\subsection{Average Treatment Effect}\label{subsec:ate}

In a nomothetic study, $Y^a_i = g^Y_a \big( U_i, \calE^Y_i \big)$ is assumed to hold for all individuals. For individual $i \in \{ 1, \hdots, n \}$ with $U_i = u$ and $\calE^Y_i = \varepsilon^Y$, this yields a set of fixed POs $\{ y^a_i \}$ with set cardinality equal to that of the support of $X$. For example, if $a \in \{ 0, 1 \}$, then individual $i$ has the two POs $y^1_i$ and $y^0_i$. The {\sl individual exposure effect} (IEE) is often defined as a contrast between $y^a_i$ and $y^{a'}_i$ for some pair $\{ a, a' \ne a \}$. While of primary interest, this IEE is not identifiable due to the fundamental problem of causal inference. However, if $X$ is randomized for all individuals, then $E \big( Y^a \big| R = 1 \big)$ (i.e., the mean PO taken across all study individuals) is identifiable using only observed data because it can be shown that

\begin{align}
E \left(
	Y
	\big|
	X = a,
	R = 1
\right)
& =
	E \left(
		Y^a
		\big|
		R = 1
	\right)
	\label{eqn:Requals1}
\end{align}
under randomization, CC, and the POGM.

We will define a {\sl treatment} to be just such a randomized exposure. A contrast of $E \big( Y^a \big| R = 1 \big)$ and $E \big( Y^{a'} \big| R = 1 \big)$ is called an {\sl average treatment effect} (ATE), and is the primary estimand of interest in nomothetic randomized studies because all other causes $U$ need not be observed in order to consistently estimate $E \big( Y^a \big| R = 1 \big)$. This result conveys the power of randomization as a tool for elucidating causal mechanisms.

If $R = 0$ for all individuals, as in an observational study, then it does not generally follow that $E \big( Y \big| X = a, R = 0 \big) = E \big( Y^a \big| R = 1 \big)$. This is because under the outcome CM and ERCM, $U$ causes both $X$ and $Y$. Hence, the measured association between $X$ and $Y$---if not accounting for $U$---generally does not reflect the true ATE (or lack thereof) of $X$ on $Y$. Here, $U$ {\sl confounds} this observed association, and so is called a {\sl confounder}.

Now suppose the randomization status of $X$ neither causes, is caused by, nor shares common causes with all other causes $U$, such that $R \indep U$. In such cases, we will say that $R$ is distributionally invariant with respect to $U$, or equivalently that {\sl distributional invariance} (DI) holds. DI implies  the penultimate equality in the statement $E ( Y^a | R ) = E_U \big\{ E ( Y^a | U, R ) \big| R \big\} = E_U \big\{ E ( Y^a | U ) \big| R \big\} = E_U \big\{ E ( Y^a | U ) \big\} = E ( Y^a )$; in particular, $E ( Y^a | R = 1 ) = E ( Y^a )$ in \eqref{eqn:Requals1}. If DI holds, $E ( Y^a )$ is still identifiable using only observed data because it can be shown that

\begin{align}
E_U \left\{
	E \left(
		Y
		\big|
		U,
		X = a,
		R = 0
	\right)
	\big|
	R = 0
\right\}
& =
	E \left(
		Y^a
	\right)
	\label{eqn:Requals0}
\end{align}
under CC. Hence, the DI condition is required for an experimental study's results to generalize or transport to a non-experimental setting (or vice versa). The expression to the left of the equals sign in \eqref{eqn:Requals0} is known variously as the g-formula, direct standardization, stratification, or regression adjustment in the causal inference literature \cite{2009_robins_hernan, 1986_robins, 1995_pearl_robins, 2006_hernan_robins, 2004_lunceford_davidian, 2014_morgan_winship}.

PO-based causal inference (at least as defined in this paper) can be understood as a special case of a general class of problems concerned with inferring causal effects. This can be seen by noting that using our CM-based definitions, the CC-POGM formulas are not needed in order to identify $E ( Y | X = a, R = 1)$ in cases where $R = 0$. Specifically, it can be shown that

\begin{align}
E_U \left\{
	E \left(
		Y
		\big|
		U,
		X = a,
		R = 0
	\right)
	\big|
	R = 0
\right\}
& =
	E \left(
		Y
		\big|
		X = a,
		R = 1
	\right)
	\label{eqn:noPOs}
\end{align}
under DI and randomization.

However, PO-based methods have been both extremely well-developed and successfully deployed for inferring the effects of {\sl manipulable} interventions or treatments. This is because PO-based designs force study investigators to explicitly define implementable interventions. For example, an outcome can (and perhaps should, to preserve information) be modeled as a function of a non-randomized exposure with continuous values (or a very large number of categories). But in order to subsequently design a corresponding randomized trial to test the assumed CM, this exposure should be transformed into a categorical treatment variable with only a few levels. This in turn limits the number of causal contrasts that are used to calculate an appropriate sample size for the trial. Such a strategy enables researchers and analysts to recommend concrete, pragmatic courses of action (e.g., treatment plans, policy changes).

Contrast this to a study that tries to infer effects comprised of a multitude of causal contrasts across a continuous-valued treatment (or one with a very large number of categories): The sample size required to achieve good statistical power to detect all possible treatment effects could be enormous due to the combinatorially large set of possible contrasts. (A plausible way to avoid needing such an exposure-to-treatment transformation might involve assuming a linearized relationship between exposure and outcome values, with the target estimand being the mean effect of a one-unit increase in the exposure.)

Practical, readily implementable interventions are our focus. Hence, we hereafter use the term \lq\lq causal inference" to specifically mean \lq\lq statistical inference of manipulable-intervention effects for a small or computationally feasible number of treatment levels."

	\subsection{Average Period Treatment Effect}\label{subsec:apte}

In an idiographic study, the individual has repeated measurements $j = 1, \hdots, m$. Let $Y^a_j$ represent the PO of $Y_j$ corresponding to $X_j = a$. The POGM $Y^a_j = g^Y_a \big( U_j, \calE^Y_j \big)$ is assumed to hold for all measurements. For measurement $j$ with $U_j = u$ and $\calE^Y_j = \varepsilon^Y$, this yields a set of fixed POs $\{ y^a_j \}$. We will call the IEE analogue in this setting a {\sl measurement-level treatment effect}, defined as a contrast between $y^a_j$ and $y^{a'}_j$ for some pair $\{ a, a' \ne a \}$, which is not identifiable due to the fundamental problem of causal inference. However, if $X_j$ is randomized for all measurements, then $E \big( Y^a_j \big)$ (i.e., the mean PO taken across all measurements) is identifiable using only observed data, by a derivation analogous to that of \eqref{eqn:Requals1}. Let $\alpha_j ( a, a' )$ denote an {\sl idiographic treatment effect}, defined as a contrast of $E \big( Y^a_j \big)$ and $E \big( Y^{a'}_j \big)$. Under analogous definitions of the POGM, CC, the ERCM, and DI, $E \big( Y^a_j \big)$ is identifiable even if $X_j$ is not randomized.

An n-of-1 study is a structured time series of outcomes, where the structure is a partition of the outcome time series imposed by a specified series of treatment periods. Let $t(j)$ denote a time point within period $t \in ( 1, \hdots, \tau )$. Each individual has repeated measurements at time points $j \in ( 1, \hdots, m_t )$ within each period $t$. In an N1RT, treatment level is randomized per period only at $t(1)$, and is otherwise kept constant. That is, we have randomized assignment $X_{t(0)} = a$ where $t(0)$ is the last time point in the previous period (i.e., $t-1(m_{t-1})$), and $X_{t(j)} = a$ for $j \in ( 1, \hdots, m_t-1 )$ if $m_t > 1$. Hence, we will write $X_t$ instead of $X_{t(j)}$ when discussing an n-of-1 study. Similarly, in an N1OS, an exposure period is an interval of time over which an exposure level is constant.

We now specify the outcome CM in terms of the POGM, and assume the relevant CC condition holds. Let $Y^a_{t(j)}$ represent the PO of $Y_{t(j)}$ corresponding to $X_t = a$. The individualized POGM $Y^a_{t(j)} = g^Y_a \big( U_{t(j)}, \calE^Y_{t(j)} \big)$ is assumed to hold across all periods and time points. For time point $j$ with $U_{t(j)} = u$ and $\calE^Y_{t(j)} = \varepsilon^Y$, this yields a set of fixed POs $\big\{ y^a_{t(j)} \big\}$. The n-of-1 measurement-level treatment effect is what we will call a {\sl point treatment effect}, defined as a contrast between $y^a_{t(j)}$ and $y^{a'}_{t(j)}$. These contrasts form a trend over a period, so we will define the {\sl period treatment effect} (PTE) as the ordered set of point treatment effects. Let $\alpha_{t(j)} ( a, a' )$ denote an {\sl average period treatment effect} at point $j$ (APTE$_j$); i.e., the corresponding idiographic treatment effect specified with $E \big( Y^a_{t(j)} \big)$ and $E \big( Y^{a'}_{t(j)} \big)$. We will refer to the period-long set of these contrasts simply as the APTE. We consider the APTE to be the primary estimand of interest in an n-of-1 study (i.e., N1RT or N1OS).

The mean PO $E \big( Y^a_{t(j)} \big)$ should be interpreted carefully as follows. Recalling \eqref{eqn:Requals1}, $E \big( Y^a_{t(j)} \big)$ equals the expected observed outcome if the subject is randomized to treatment $a$ in period $t$---but not over all periods, as would be directly analogous to the interpretation of the mean PO of an ATE (i.e., taken over all individuals). This is an important distinction, because randomization to $a$ at all time points would violate the DI condition.

N-of-1 studies commonly involve the following outcome-related complications. Outcomes may be {\sl autocorrelated} such that $Y_{t(j)}$ depends on a set of lagged outcomes $\bar{Y}_{t(j')} = \big\{ Y_{t(j')} \big\}_{\{ j' < j \}}$, where some $j'$ may index time points in previous periods $\{ t' \}_{\{ t' < t \}}$. We write $Y_{t(j)} = g^Y \big( \bar{Y}_{t(j')}, \calE^Y_{t(j)} \big)$ if all of these lagged outcomes cause the current outcome to change. There may be a {\sl time trend} over periods in the outcomes such that $E \big( Y_{t(j)} \big)$ is not constant at each time point regardless of period; e.g., $E \big( Y_{t(j)} \big)$ increases as $t$ increases.

N-of-1 studies also commonly involve the following treatment-related complication. At least one PO $Y^a_{t(j)}$ may depend on a set of lagged treatments $\bar{X}_{t'} = \big\{ X_{t'} \big\}_{\{ t' < t \}}$. We write $Y^a_{t(j)} = g^Y_a \big( \bar{X}_{t'}, \calE^Y_{t(j)} \big)$ if all of these lagged treatments cause the current PO to change. In addition, if the treatment effect $\alpha_{t(j)} ( a, a' )$ itself depends on $\bar{X}_{t'}$, we say that the treatment (specifically, the treatment level) {\sl carries over} to the current period from past periods. For example, let $a \in \{ 0, 1 \}$ and write $\alpha_{t(j)}$ instead of $\alpha_{t(j)} ( 1, 0 )$ as shorthand, and suppose $E \big( Y^0_{t(j)} \big) = 1$ always. Suppose $E \big( Y^1_{t(j)} \big| X_{t-1} = 0 \big) = 6$ at all $j$, but $E \big( Y^1_{t(j)} \big| X_{t-1} = 1 \big) = 2$ at all $j$. Then $\alpha_{t(j)} = 5$ at all $j$ in the first case, but $\alpha_{t(j)} = 1$ at all $j$ in the second case. The term \lq\lq carryover effect" is sometimes used, where \lq\lq effect" does not mean a causal effect as we have defined it. (The closest definition of a carryover \lq\lq effect" might be the difference in the treatment effect resulting from carryover compared to when no carryover is present. In particular, we do not define what it means for the treatment effect itself to carry over.) We henceforth say \lq\lq {\sl treatment carryover}" (or simply \lq\lq {\sl carryover}") instead of \lq\lq carryover effect" to avoid confusion.

Carryover may induce a within-period time trend in the treatment effect such that $\alpha_{t(j)} ( a, a' )$ changes over the period. For example, let $m_t = 3$ for all $t$, and suppose as before that $E \big( Y^0_{t(j)} \big) = 1$ always. Suppose $E \big( Y^1_{t(j)} \big| X_{t-1} = 0, X_{t-2} = 0 \big) = 6$ at all $j$, $E \big( Y^1_{t(j)} \big| X_{t-1} = 1, X_{t-2} = 0 \big) = 6 / j$, and $E \big( Y^1_{t(j)} \big| X_{t-1} = 1, X_{t-2} = 1 \big) = 2$ at all $j$. Then $\alpha_{t(j)} = 5$ at all $j$ in the first case, $\big( \alpha_{t(1)}, \alpha_{t(2)}, \alpha_{t(3)} \big) = ( 5, 2, 1 )$ in the second case, and $\alpha_{t(j)} = 1$ at all $j$ in the third case. The second case is an example of {\sl slow decay} of the treatment effect, with the opposite trend called {\sl slow onset}.

In an N1RT, randomization eliminates confounding due to autocorrelation and carryover. There is no such guarantee in an N1OS; the possibility of confounding cannot be ignored. Thankfully, a number of methods exist to address or adjust for autocorrelation- and carryover-induced confounding. These are detailed in the next section.

	\subsection{Modeling and Estimating the APTE}\label{subsec:apteestimation}

The ATE literature provides a number of key techniques for modeling and estimating the APTE when exposure is not randomized. These either model the outcome or PO, the exposure, or both; some methods model the causal contrast itself. Perhaps the most straightforward approach is to first model the outcome as a function of the exposure and other causes, and then take the expectation over all other causes at a given exposure level to yield the mean PO for that level without conditioning on the exposure (i.e., as if exposure had been randomized). The general expression for this {\sl g-formula} approach is \eqref{eqn:Requals0}, with its corresponding {\sl outcome model}. For categorical or binary exposures, other popular approaches model the exposure probability as a {\sl propensity score}; i.e., the probability of receiving an exposure level, which might depend on other causes \cite{1983_rosenbaum_rubin}. These approaches adjust the observed outcomes such that they are a function of the actual exposure received and the propensity score; e.g., through matching or inverse probability weighting (IPW). These models are called {\sl exposure} or {\sl propensity models}. Advanced techniques that can improve statistical efficiency model both the outcome and exposure CMs (i.e., \lq\lq augmented IPW" or \lq\lq doubly robust estimation") that only require one of the two models to be correctly specified to guarantee statistically consistent estimation of the ATE. Daza (2018) \cite{2018_daza} applied both the g-formula and IPW methods to estimate APTEs on continuous outcomes for a binary-valued exposure, by specifying models that allowed for autocorrelation, carryover, and slow onset.

Suppose neither the outcome nor exposure model is known, or cannot otherwise be reasonably justified (e.g., due to a lack accumulated theoretical evidence). This is the case in our setting of causal hypothesis generation: Our primary objective is to {\sl identify} or {\sl select} plausible models for a small set of posited causal effects, rather than {\sl estimate} effects and associations posited by scientifically defensible or otherwise well-established a priori models \cite{2010_arlot_celisse, 2007_yang}. Here, we use supervised learning methods that allow us to fit models focused on the exposure of interest while flexibly accommodating non-exposure causes. The decision tree method is one such non-parametric approach, that Athey and Imbens (2015)\cite{2015_athey_imbens} used to estimate a conditional ATE for a continuous outcome (specified as a difference between mean POs). Their tree-based approach can be used to estimate an APTE if conditional wide-sense stationarity (WSS) holds (i.e., the outcomes are WSS conditional on the causes); WSS replaces the assumption of conditional independence (i.e., the outcomes are mutually independent conditional on the causes) needed for consistent effect estimation. For example, the Two Trees approach \cite{2015_athey_imbens} for conducting feature selection (i.e., for non-exposure causes) separately models the POGM corresponding to each exposure level as a function of the non-exposure causes. Following Athey and Imbens (2015)\cite{2015_athey_imbens}, we took a Single Tree approach, and modeled the outcome as a function of the exposure and other causes using random forests (RF).

At each time point, we consider the possibility of observing four types of non-exposure causes: Confounders, {\sl simultaneous causes} (i.e., occurring simultaneously with the exposure), {\sl mediators} (i.e., follow the exposure and are themselves caused by the exposure), and {\sl post-exposure causes} (i.e., follow but are not caused by the exposure). All other predictors or features that may be conditioned on in specifying a treatment effect that are not well-defined (i.e., effect {\sl moderators}) are assumed for simplicity to be one of these four causal types, depending on their temporal relationship to the exposure and outcome.

We estimate each mean PO by first predicting the PO conditional on the exposure and other causes. We subsequently take the average (i.e., empirical mean) over a subset of these causes using the g-formula. Performing this marginalization correctly requires knowing which variables are confounders and simultaneous causes, mediators, and post-exposure causes in temporal relation to the outcome and exposure. In our current approach we do not condition on the exposure when marginalizing over confounders or simultaneous causes, but do condition on the exposure when marginalizing over mediators and post-exposure causes. This approach produces estimates of the average {\sl total effect}; i.e., the total of the {\sl direct effect} of the exposure alone, and the {\sl indirect effects} of the exposure through mediators. Importantly, the role of each timepoint-length (i.e., as opposed to period-length) variable can change at different time points: A confounder at one time point may be a simultaneous cause, mediator, or post-exposure cause at a subsequent time point. This follows because we fit the same outcome model at all time points, while the time from the introduction of the treatment increases.

Each confounder or simultaneous cause (CSC) must be stationary for the APTE g-formula estimator to be consistent for an effect-stable APTE. Stationarity of each CSC can be formally tested using the Augmented Dickey-Fuller (ADF) and Kwiatkowski-Phillips-Schmidt-Shin (KPSS) unit-root tests, implemented via the respecdtive R commands {\tt adf.test()} and {\tt kpss.test()}. Let $S_{j p_j} = 1$ if CSC $p_j$ at time point $j$ passed either test, and let $S_{j p_j} = 0$ otherwise. As a rough way to assess the statistical consistency of our findings for an effect-stable APTE, we define the overall CSC stationarity as the empirical mean of all $S_{j p_j}$ taken over all CSCs at all time points.

	\subsection{Visualizing the APTE}\label{subsec:pancit}

To help visualize observations, along with the trajectories of corresponding predictions and estimated mean POs, we introduce the {\sl pancit plot}. First recall a spaghetti plot, which is a graph of outcome trajectories corresponding to distinct subjects or individuals in a longitudinal analysis. In an n-of-1 study, we have a time series that is partitioned into distinct treatment periods. The plot of overlaid outcome trajectories (each corresponding to a distinct period) resembles a spaghetti plot. Hence, we will call such a plot a pancit plot, from the word \lq\lq pancit"---pronounced like \lq\lq pun-SEAT", rhymes with \lq\lq receipt"---meaning \lq\lq noodles" in Filipino.

Note that unlike a longitudinal study, we predict each outcome as a function of preceding variables. Hence, while a spaghetti plot only illustrates observed outcomes, a pancit plot depicts both observed and predicted outcomes; in our case, predicted POs (PPOs). We draw observed outcomes as dots (i.e., the \lq\lq sauce"), and draw PPOs as lines (i.e., the \lq\lq noodles").

	\subsection{Analysis Procedure}\label{subsec:analysisproc}

To discover, specify, and estimate APTEs, we based our procedure on that of Daza (2018)\cite{2018_daza} as follows.

\begin{enumerate}

	\item {\sl Define the outcome.} If possible, identify stationary segments of the outcome; the analysis should be conducted separately over each segment. This may help justify the assumption that the estimated APTE is {\sl effect-stable} (i.e., the APTE is constant across periods \cite{2018_daza}). One way to do this is changepoint detection, in which an algorithm detects the points in a stationary time series at which the mean value changes \cite{2014_killick_eckley}.

	\item {\sl Define the periods, treatment, and APTE.} If not already identified, use a reasonable technique to specify periods. For example, perform changepoint detection on the continuous exposures to construct plausible periods over which the exposure level remains constant. Once exposure periods have been identified, transform the exposure into a binary-valued treatment by setting a threshold to indicate high (i.e., at or above the threshold) or low (i.e., below the threshold) exposure as follows: Perform predictive modeling for each threshold for a set of thresholds. Then select the threshold resulting in the largest prediction success metric. Specify the APTE set of contrasts (e.g., differences, ratios) over a period.

	\item {\sl Define the non-exposure causes.} These can include lagged outcomes and exposures. Note that estimating the mean POs needed to calculate the APTE involves averaging the predicted outcomes over the set of non-exposure causes. If there are too many non-exposure causes, constructing their empirical joint distribution (i.e., to be used in marginalization) can quickly become intractable. In addition, including too many non-exposure causes in a predictive model may bias the APTE estimates by inducing a type of spurious association called M-bias \cite{2009_pearl}. To address this, reduce the set of non-exposure causes; e.g., by selecting the most important predictors through a parameter- or model-selection procedure.

	\item {\sl Estimate the APTE.} First predict the PO for each set of observed values (i.e., exposure and other causes) at each time point. Then take the weighted average over all non-exposure causes at each time point to produce the trajectory of estimated mean POs for each exposure level, with weights derived from the empirical marginal CSC distribution as required by the g-formula. Report the estimated APTE as the trajectory of pre-defined contrasts. If desired, report the naively estimated mean POs and APTE by taking the simple rather than weighted average (i.e., conditional on treatment level at each time point) in the aforementioned procedure.

	\item {\sl Produce plots and numerical summaries.} Plot the time series of observed outcomes and PPOs, create a corresponding pancit plot, and plot the estimated APTE. Note that the APTE contrast at a given time point reflects the total effect at that time point. If relevant, also create the pancit and APTE plots using naive estimates.

	\item {\sl Interpret the findings.} Discuss the scientific meaning of the estimated APTE and its implications. Take care to address missing data and other phenomena that might have affected results. Perform sensitivity anaylses if warranted. Point out limitations, and suggest future improvements or directions.

\end{enumerate}
Because our outcome was continuous, we chose the test/generalization mean squared error (MSE) as our success metric, estimated using out-of-bag cross-validation performed by RF \cite{2001_friedman_etal}. The model chosen therefore produced the lowest estimated test MSE.

\section{Empirical Application}\label{sec:emp.example}

We analyzed the dataset referenced in Daza (2018)\cite{2018_daza} spanning six years of the author's body weight and physical activity (PA). As in that paper, we hypothesized that high PA causes weight loss, while low PA causes weight gain. We used a g-formula approach, and hence specified the outcome CM as detailed below. All hypothesis tests were performed at the 0.05 significance level unless stated otherwise. All variables used in our analysis consisted of outcomes, exposures, and their lags. We executed the procedure outlined in Section \ref{subsec:analysisproc} as follows.

	\subsection{Define the outcome.}

The analysis outcome was defined as average centered body weight (ACBW) per week, centered around the empirical average of body weight measured in kilograms over the six-year study period. The analysis exposure was defined as the proportion of days per week with any PA reported (among days with non-missing body weight). PA on any given day consisted of cardiovascular or resistance training, in line with usual definitions in the literature \cite{2012_trinh_etal, 2015_naimark_etal, 2016_afshin_etal, 2016_schoeppe_etal}. This yielded a time series of 290 time points. Of these, the ACBW was considered to be stationary using changepoint detection for 211 consecutive time points using the {\tt changepoint::cpt.mean()} R command\cite{2014_killick_eckley, 2016_killick_etal}. We used the latter as our analysis dataset in order to simplify exposition by avoiding a possibly strong time trend.

	\subsection{Define the periods, treatment, and APTE. Define the non-exposure causes.}

We defined treatment period as follows. First, we defined the exposure as a regular or stable level of PA observed over at least 11 weeks, based on our findings in Daza (2018)\cite{2018_daza}. This resembles the length of previous exposure or treatment periods for observing the effects of similar interventions, which range from six to 14 weeks \cite{2016_schoeppe_etal, 2016_afshin_etal}. In Daza (2018)\cite{2018_daza}, we used changepoint detection to partition the series into segments of constant mean exposure. We then considered the segment lengths and means to be fixed, and equated the segments with putative treatment periods. We determined the cutoff value with which to dichotomize exposure into treatments later on below.

We first reduced the set of non-exposure causes as follows. For simplicity in exposition, we set the following somewhat arbitrary parameters with the intent to account for confounding by lagged variables, while not specifying so many lags so as to reduce the effective sample size (i.e., by requiring each observation to have a large number of corresponding lagged observations). Hence, we defined the initial putative non-exposure causes as consisting of up to three outcome lags and 12 exposure lags. To identify the most important predictors, we conducted RF using the {\tt randomForest::randomForest()} R function \cite{2002_liaw_wiener}; the 18 predictors consisted of exposure, these lagged variables, time-ordered period number, and within-period week number.

We first defined the putative treatment as follows. We fit RF models with the selected predictors at the three exposure thresholds corresponding to the 25th, median, and 75th quantiles of the empirical distribution of segment means. We found that the median produced the model with the smallest average MSE. Hence, treatment was defined as high PA if the mean exposure for a period was at or above the median, and as low PA otherwise. The APTE was defined as the trend in differences in mean ACBW POs.

To keep the analysis tractable, we chose to limit the number of CSCs over which to marginalize. Specifically, we kept the nine most important predictors, defined as having the largest mean decreases in MSE \cite{2002_liaw_wiener}. (A more structured approach is briefly mentioned in the Section \ref{sec:discuss}.) In order of decreasing importance, these were outcome lags 1 and 2, week (in period), outcome lags 3 and 4, period, and outcome lags 6, 12, and 7. Treatment itself was only ranked 15th in importance. The reduced model of selected predictors consisted of these nine predictors and treatment, which we included for reasons stated below. The out-of-bag MSE was 0.27.

Three findings warrant further discussion. We interpreted the predictive period number as indicating a time trend. Similarly, we interpreted the predictive week number as indicating time-dependent autocorrelation effects or carryover within a period (e.g., analogous to the interpretation of interaction terms in a linearized model). Week number could also function as a proxy variable for latent, unobserved confounders with time-dependent effects within period \cite{2018_daza}. Lastly, even though treatment was not ranked as highly important, we included it in subsequent model fitting because our a priori goal was to estimate its effect on the outcome; all other predictors would be used to improve effect estimation. That is, we wanted to perform model identification/selection for all other predictors \cite{2010_arlot_celisse, 2007_yang} under the assumption that a true treatment effect existed.

	\subsection{Estimate the APTE. Produce plots and numerical summaries.}

The main APTE findings are as follows. Period length ranged from 1 to 44 weeks, with a median of 5 weeks. There were 33 low-PA periods and nine high-PA periods. These can be seen in Figure \ref{fig1_ts}, a time series of observed ACBW outcomes and corresponding PPOs for each treatment period; periods are separated by vertical lines. For example, period 8 was the third high-PA period, with observed ACBW of -0.93 and PPO of -0.52 at week 1. (It was also the longest period, with 44 total weeks.) Figure \ref{fig2_pancit} depicts these same data, but as a pancit plot. For example, the estimated mean POs at week 1 under low and high PA were -0.264 and -0.406, respectively. As can be seen, the longest timespan over which we were able to estimate the mean PO under both low and high PA was 12 weeks. Hence, the APTE was estimable only over 12 weeks, and is illustrated in Figure \ref{fig3_apte}. For example, the estimated APTE at week 1 (i.e., APTE$_1$) was an average gain of 0.143 kg. Table \ref{tab1_PPOsAPTEs} lists all estimated mean POs (i.e., average PPOs) and the corresponding estimated APTE. The overall CSC stationarity in our analysis was 0.929.

The naively estimated mean POs and APTE closely resemble their correct counterparts. This can clearly be seen in Figures \ref{fig2_pancit} and \ref{fig3_apte}, and in Table \ref{tab1_PPOsAPTEs}.

	\subsection{Interpret the findings.}

The estimated APTE suggests, perhaps contradictory to expectation, that high PA initially results in average weight gain during the first five weeks of a high-PA period. This can be seen in Figure \ref{fig3_apte} and Table \ref{tab1_PPOsAPTEs} to indicate an average weight gain of about 0.2 kg in the first two weeks. However, this initial weight gain changes to weight loss after five weeks; within 12 weeks, this average weight loss seems to stabilize around -0.3 kg sometime after seven weeks. The high overall CSC stationarity indicates that the APTE we estimated is likely effect-stable (if our estimator is statistically consistent). Hence, the estimated APTE is likely to persist unchanged over time; at least, across similar distributions of CSC values.

The close resemblance of the naive estimates to the correct estimates highlights an important limitation of our study. At first, this similarity in estimates might indicate a lack of confounding. Specifically, per our findings, suppose the author's current body weight can only be affected by body weight as far back as 12 weeks, and by PA level during the current period. If past body weight does not influence the author's PA tendencies, then the naive and correct mean POs and APTE would be identical, as their estimates in Figures \ref{fig2_pancit} and \ref{fig3_apte} seem to indicate. However, diet is known to impact weight trends. The similarity in naive and correct estimates may therefore be a result of excluding this potentially strong confounder in our models, rather than there being a lack of confounding in general. For example, consistent consumption of rich or heavy meals (as around a holiday period) may have both increased the author's body weight and decreased his tendency to exercise during a subsequent period. A future analysis using these data should therefore account for the author's dietary patterns as well, if possible.

To help answer some of our immediate questions, a follow-up study could take a propensity-score approach. The counterintuitive shift in APTE from weight gain to weight loss may be explained by noting in Figure \ref{fig2_pancit} that low-PA periods typically start out with lower body weights. This could make sense if low-PA periods typically follow high-PA periods, as is done in propensity modeling. Such a propensity model could also be fit using lagged body weight in addition to lagged PA levels, to see how past body weight might affect the propensity of the author to engage in high or low PA. However, this approach will likely require more than nine high-PA periods, as per our current study.

Future studies should also explore the sensitivity analyses implemented or suggested in Daza (2018) \cite{2018_daza}, which include the following. The original dataset included body weight and PA recorded daily, from which ACBW and weekly PA proportion was calculated. However, many of these raw outcomes and exposures were missing; hence, some of the resulting weekly outcomes and exposures were also missing (i.e., if no body weights or PA indicators were recorded for an entire week). Our analysis dataset was constructed by assuming, for example, that the trend in weekly PA proportion was linear in cases when they were missing, and were otherwise missing completely at random \cite{1976_rubin, 2014_little_rubin}. The missing data points were imputed using the {\tt na.interpolation()} command. We also used the {\tt Amelia::amelia()} command \cite{2011_honaker_etal} to impute missing ACBW values. Future studies should examine the sensitivity of our findings to such simplifying assumptions (which could also have affected the changepoint-identified exposure segments used to define treatment periods). In addition, Daza (2018) varied the start day used to define each week over which to calculate the ACBW and weekly PA proportion. Future work should likewise examine how (if at all) our findings change as the start day is varied.

\section{Discussion}\label{sec:discuss}

In summary, we have defined an individualized causal effect, the APTE, and have argued that it can be understood as a type of longitudinal ATE. We first used Daza (2018)\cite{2018_daza} to define counterfactual-based causal inference concepts used to define and model the APTE. We then analyzed six years of the author's self-tracked body weight and PA data using a g-formula approach based on random-forests modeling. We found that over a 12-week period, high PA may have caused the author to first gain about 0.2 kg, but then start losing weight around five weeks, with average weight loss possibly stabilizing to -0.3 kg after about seven weeks. However, the lack of other notable confounders such as diet in our models may have biased our APTE estimates. We also suggested that a propensity-score approach be taken in future studies, and proposed some relevant sensitivity analyses to missing data and outcome/exposure start day.

How might such an idiographic ATE relate to a corresponding nomothetic ATE? One common approach is to combine subjects' estimated ATEs in the same way ATEs from multiple nomothetic studies would be aggregated in a meta-analysis. However, if an APTE is a longitudinal ATE for a single subject, then another natural approach appears by noting that a particular subject's outcomes may be clustered together with respect to other individuals' outcomes. Hence, another way to aggregate multiple subjects' data is to model the APTE {\sl clustered by individual}, as in a mixed-effects model in a longitudinal study. Zucker et al (1997, 2010)\cite{1997_zucker_etal, 2010_zucker_etal} examine these and other approaches in detail for point treatment effects, and their work helps build a foundation with which to extend our single-subject approach for period treatment effects.

The following points from Daza (2018)\cite{2018_daza} are worth re-iterating. In the APTE framework, each time point $t(j)$ can contain sub-points (e.g., $t(j_k)$). This modularity allows for flexible temporal scaling of posited causal relationships. The framework's utility can also be complemented and improved with formal causal diagrams such as directed acyclic graphs (DAGs) \cite{1995_pearl, 2007_vanderweele_robins} for conceptualizing formal structures. We are currently developing the theory needed to connect DAGs to the APTE framework.

We are happy to report that work is already underway to deepen such N1OS causal approaches in both theoretical and applied directions. Van der Laan and Malenica (2018)\cite{2018_vanderlaan_malenica} have been developing an approach to estimate APTEs for point treatments using a targeted maximum likelihood estimation (TMLE). Their important paper lays the theoretical groundwork for both the asymptotic consistency and normality of their TMLE estimator, and proposes a sequentially adaptive design for learning the optimal idiographic treatment rule over time. We ourselves are completing work on a forthcoming R package, tentatively titled {\tt CAPTEuR} (Calculating the APTE using R), to implement the procedure in Section \ref{subsec:analysisproc}. N-of-1 approaches are increasing in popularity due to the availability of relevant idiographic measurements. It is exciting to be able to contribute to the rigorous causal analysis of such personally meaningful data.

\section{Acknowledgements}

This work was supported by the Stanford Clinical and Translational Science Award (CTSA) to Spectrum (UL1 TR001085). The CTSA program is led by the National Center for Advancing Translational Sciences (NCATS) at the National Institutes of Health (NIH). The content is solely the responsibility of the author and does not necessarily represent the official views of the NIH. This research was also supported by the National Institutes of Health (NIH) grant 2T32HL007034-41. The content is solely the responsibility of the author and does not necessarily represent the official views of the NIH. I thank Prof. Michael Baiocchi for his constant support, feedback, and guidance. Thank you to my family and friends for their constant support. In addition, I thank my colleagues, adversaries, teachers, students, and most of all, luck and privilege. I dedicate this paper to Filipinos and Filipino-Americans. To all who are underrepresented or unacknowledged in science, technology, engineering, and math (STEM), and in academia: Kaya natin 'to! To you, the reader: Know yourself, help others, and find meaning in all things.

\bibliographystyle{vancouver}
\bibliography{paplong_refs}


\section*{}

\afterpage{

\begin{figure}
\begin{center}
\includegraphics[scale=0.66]{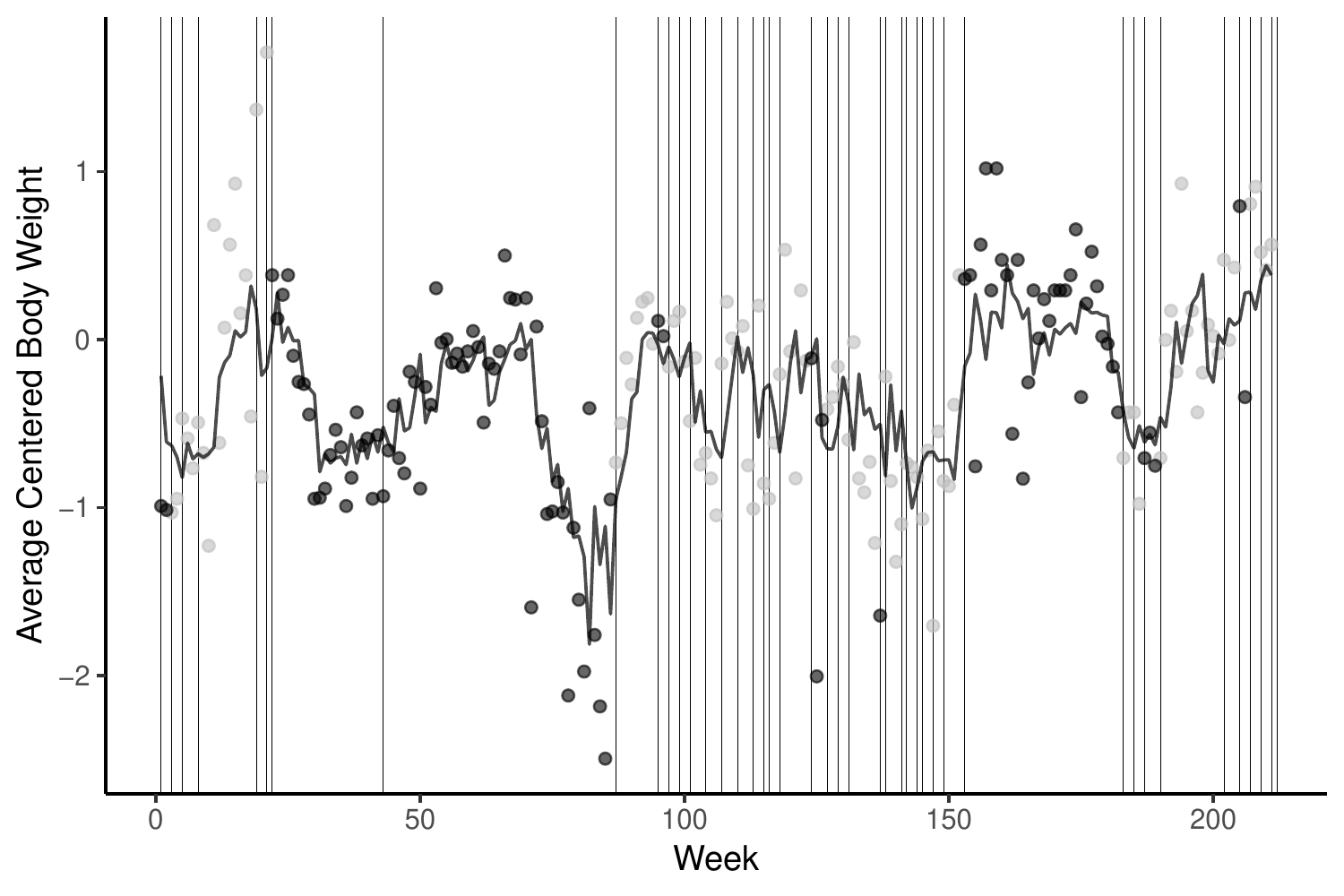}
\end{center}
\caption{ \label{fig1_ts}Time series plot of observed outcomes (dots) and predicted potential outcomes (line). Black (gray) dots represent average centered body weight during periods of high (low) physical activity. Periods are separated by vertical gray lines.}
\end{figure}

\begin{figure}
\begin{center}
\includegraphics[scale=0.66]{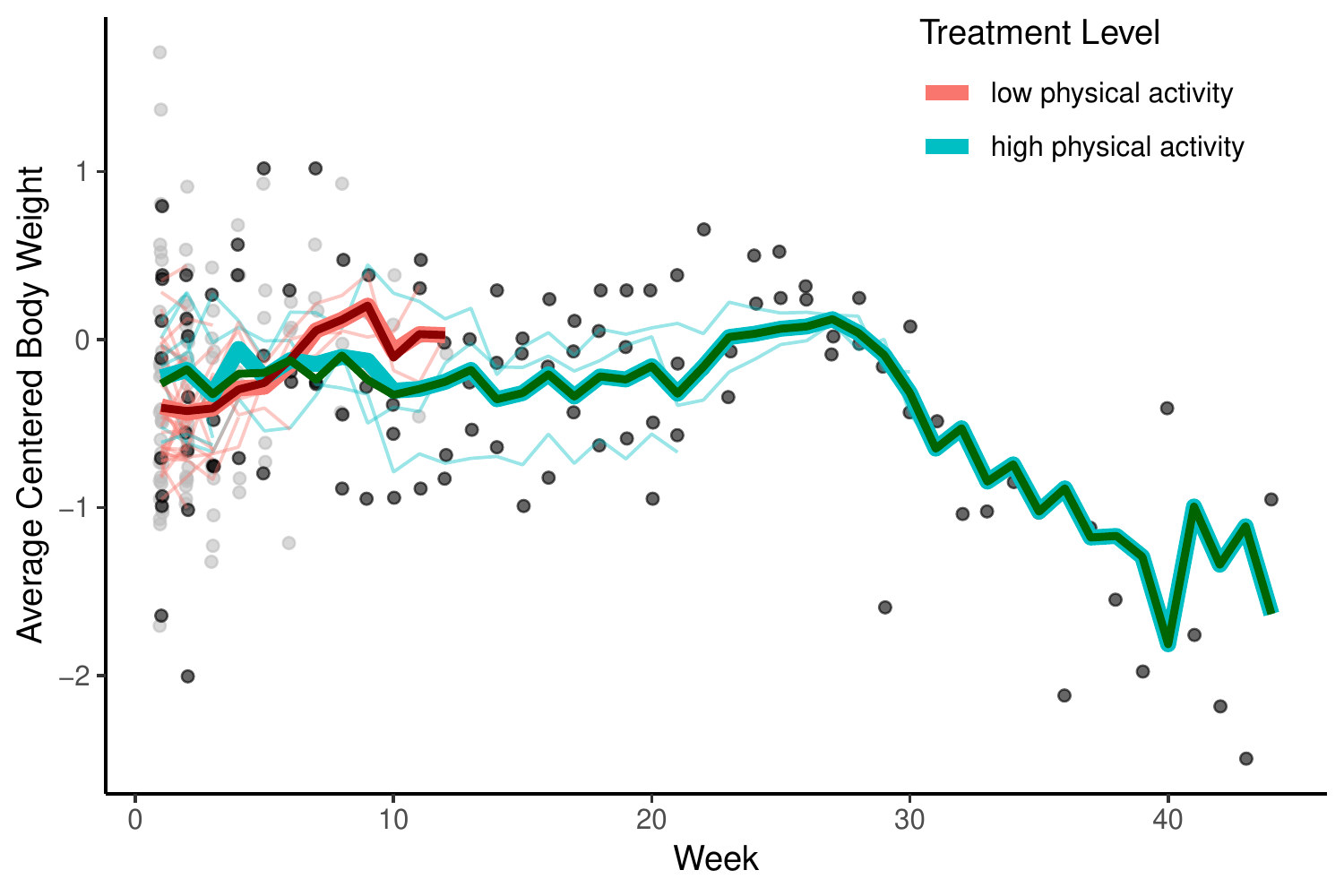}
\end{center}
\caption{ \label{fig2_pancit}Pancit plot of observed outcomes (dots) and predicted potential outcomes (POs) (noodles). Black (gray) dots represent average centered body weight during periods of high (low) physical activity (PA). Teal (salmon) small noodles represent predicted POs under high (low) PA. Thinner/darker (wider/lighter) big noodles represent mean POs estimated correctly (naively).}
\end{figure}

\begin{figure}
\begin{center}
\includegraphics[scale=0.66]{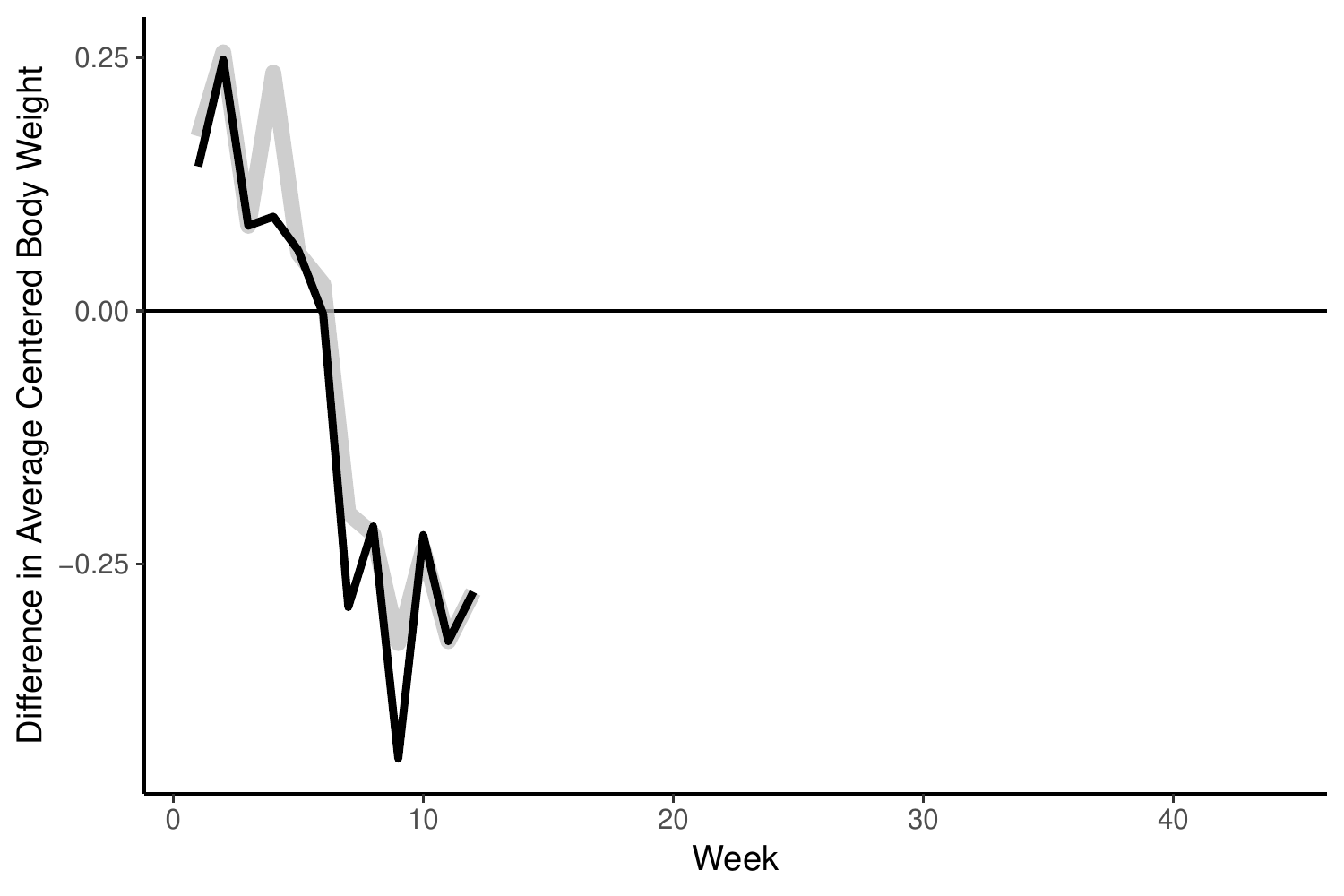}
\end{center}
\caption{ \label{fig3_apte}Average period treatment effect of physical activity level on the difference in mean average centered body weight. The thinner/darker/black (wider/lighter/gray) bold noodle represents the APTE estimated correctly (naively).}
\end{figure}

\setlength{\extrarowheight}{4pt}
\begin{table}
\caption{ \label{tab1_PPOsAPTEs} Estimated mean predicted potential outcomes and APTE, with corresponding naive estimates.}
\begin{center}
\scalebox{0.77}{
\begin{tabular}{ l | rrr | rrr }
\hline
Week &$\hat{E} \left( Y^1 \right)$ & $\hat{E} \left( Y^0 \right)$ & $\widehat{APTE}$ & $\hat{E} \left( Y^1 \right)$ (Naive) & $\hat{E} \left( Y^0 \right)$ (Naive) & $\widehat{APTE}$ (Naive) \\
\hline
1 &-0.264 & -0.406 & 0.143 & -0.225 & -0.397 & 0.172 \\
2 &-0.177 & -0.425 & 0.248 & -0.181 & -0.437 & 0.255 \\
3 &-0.326 & -0.41 & 0.084 & -0.326 & -0.41 & 0.084 \\
4 &-0.204 & -0.297 & 0.093 & -0.056 & -0.291 & 0.235 \\
5 &-0.198 & -0.258 & 0.06 & -0.224 & -0.281 & 0.057 \\
6 &-0.123 & -0.119 & -0.003 & -0.123 & -0.149 & 0.026 \\
7 &-0.239 & 0.054 & -0.293 & -0.146 & 0.054 & -0.2 \\
8 &-0.095 & 0.118 & -0.213 & -0.103 & 0.118 & -0.221 \\
9 &-0.24 & 0.202 & -0.442 & -0.126 & 0.202 & -0.328 \\
10 &-0.329 & -0.108 & -0.221 & -0.305 & -0.07 & -0.235 \\
11 &-0.294 & 0.032 & -0.326 & -0.294 & 0.032 & -0.326 \\
12 &-0.251 & 0.026 & -0.278 & -0.251 & 0.026 & -0.278 \\
13 &-0.181 &  &  & -0.181 &  &  \\
14 &-0.356 &  &  & -0.356 &  &  \\
15 &-0.319 &  &  & -0.319 &  &  \\
16 &-0.207 &  &  & -0.207 &  &  \\
17 &-0.339 &  &  & -0.339 &  &  \\
18 &-0.22 &  &  & -0.22 &  &  \\
19 &-0.238 &  &  & -0.238 &  &  \\
20 &-0.159 &  &  & -0.159 &  &  \\
21 &-0.322 &  &  & -0.322 &  &  \\
22 &-0.163 &  &  & -0.163 &  &  \\
23 &0.015 &  &  & 0.015 &  &  \\
24 &0.034 &  &  & 0.034 &  &  \\
25 &0.064 &  &  & 0.064 &  &  \\
26 &0.077 &  &  & 0.077 &  &  \\
27 &0.121 &  &  & 0.121 &  &  \\
28 &0.04 &  &  & 0.04 &  &  \\
29 &-0.092 &  &  & -0.092 &  &  \\
30 &-0.318 &  &  & -0.318 &  &  \\
31 &-0.649 &  &  & -0.649 &  &  \\
32 &-0.528 &  &  & -0.528 &  &  \\
33 &-0.845 &  &  & -0.845 &  &  \\
34 &-0.743 &  &  & -0.743 &  &  \\
35 &-1.024 &  &  & -1.024 &  &  \\
36 &-0.886 &  &  & -0.886 &  &  \\
37 &-1.178 &  &  & -1.178 &  &  \\
38 &-1.169 &  &  & -1.169 &  &  \\
39 &-1.293 &  &  & -1.293 &  &  \\
40 &-1.813 &  &  & -1.813 &  &  \\
41 &-0.993 &  &  & -0.993 &  &  \\
42 &-1.34 &  &  & -1.34 &  &  \\
43 &-1.111 &  &  & -1.111 &  &  \\
44 &-1.633 &  &  & -1.633 &  &  \\
\hline
\end{tabular}
}
\end{center}
\end{table}
\setlength{\extrarowheight}{0pt}


}

\end{document}